\documentclass[aps,showpacs,twocolumn]{revtex4}

\usepackage{graphicx}
\usepackage{epstopdf}
\usepackage{dcolumn}

\begin{document}
\title{Correlating radii and electric monopole transitions of atomic nuclei}
\author{
S.~Zerguine$^{1,2}$,
P.~Van~Isacker$^2$,
A.~Bouldjedri$^1$,
and
S.~Heinze$^3$}
\affiliation{
$^1$Department of Physics, PRIMALAB Laboratory,
University of Batna, Avenue Boukhelouf M El Hadi,
05000 Batna, Algeria}
\affiliation{
$^2$Grand Acc\'el\'erateur National d'Ions Lourds,
CEA/DSM--CNRS/IN2P3, B.P.~55027, F-14076 Caen Cedex 5, France}
\affiliation{
$^3$Institute of Nuclear Physics, University of Cologne,
Z\"ulpicherstrasse 77, 50937 Cologne, Germany}
\date{\today}

\begin{abstract}
A systematic analysis of the spherical-to-deformed shape phase transition
in even-even rare-earth nuclei from $_{58}$Ce to $_{74}$W
is carried out in the framework of the interacting boson model.
These results are then used to calculate
nuclear radii and electric monopole (E0) transitions with the same effective operator.
The influence of the hexadecapole degree of freedom ($g$ boson)
on the correlation between radii and E0 transitions thus established,
is discussed.
\end{abstract}
\maketitle

Electric monopole (E0) transitions between nuclear levels
proceed mainly by internal conversion
with no transfer of angular momentum to the ejected electron.
For transition energies greater than $2m_{\rm e}c^2$,
electron-positron pair creation is also possible;
two-photon emission is possible at all energies but extremely improbable.
The total probability for a transition between initial and final states
$|{\rm i}\rangle$ and $|{\rm f}\rangle$
can be separated into an electronic and a nuclear factor, $P=\Omega\rho^2$,
where the nuclear factor $\rho$ is~\cite{Church56}
\begin{equation}
\rho=\sum_{p=1}^Z
\langle{\rm f}|
\left(\frac{r_p}{R}\right)^2
-\sigma\left(\frac{r_p}{R}\right)^4+\cdots
|{\rm i}\rangle,
\label{e_rho}
\end{equation}
with $R=r_0A^{1/3}$ ($r_0=1.2$~fm)
and where the summation runs over the $Z$ protons in the nucleus.
The coefficient $\sigma$ depends on the assumed nuclear charge distribution
but in any reasonable case it is smaller than 0.1
and can be neglected if the leading term is not too small~\cite{Church56}.

The charge radius of a state $|{\rm s}\rangle$ is given by
\begin{equation}
\langle r^2 \rangle_{\rm s}=
\frac{1}{Z}\langle{\rm s}|\sum_{p=1}^Zr^2_p|{\rm s}\rangle.
\label{e_r2a}
\end{equation}
It is found experimentally that
the addition of neutrons produces a change in the nuclear charge distribution,
an effect which can be parametrized
by means of neutron and proton effective charges $e_{\rm n}$ and $e_{\rm p}$
in the charge radius operator $\hat T(r^2)$.
This leads to the following generalization of Eq.~(\ref{e_r2a}): 
\begin{equation}
\langle r^2 \rangle_{\rm s}\equiv
\langle{\rm s}|\hat T(r^2)|{\rm s}\rangle=
\frac{1}{e_{\rm n}N+e_{\rm p}Z}
\langle{\rm s}|\sum_{k=1}^Ae_kr^2_k|{\rm s}\rangle,
\label{e_r2}
\end{equation}
where the sum is over all nucleons
and $e_k=e_{\rm n}$~($e_{\rm p}$) if $k$ is a neutron (proton).

An obvious connection between $\rho$
and the nuclear charge radius
is established in the approximation $\sigma=0$
(which henceforth will be made).
Again because of the polarization effect of the neutrons,
one introduces an E0 operator of the form~\cite{Kantele84}
\begin{equation}
\hat T({\rm E0})=\sum_{k=1}^Ae_kr^2_k.
\label{e_e0}
\end{equation}
The $\rho$ defined in Eq.~(\ref{e_rho}) with $\sigma=0$
is then given by
$\rho=\langle{\rm f}|\hat T({\rm E0})|{\rm i}\rangle/eR^2$.
The basic hypothesis of this Letter
is to assume that {\em the effective nucleon charges
in the charge radius and E0 transition operators are the same}.
If this is so, comparison of Eqs.~(\ref{e_r2}) and~(\ref{e_e0})
leads to the relation
\begin{equation}
\hat T({\rm E0})=(e_{\rm n}N+e_{\rm p}Z)\hat T(r^2).
\label{e_e0r2}
\end{equation}

At present, a quantitative test of the correlations
between radii and E0 transitions implied by~(\ref{e_e0r2})
cannot be obtained in the context of the nuclear shell model.
The main reason is that E0 transitions
between states in a single harmonic-oscillator shell
vanish identically~\cite{Wood99}
and a non-zero E0 matrix element is obtained only if valence nucleons
are allowed to occupy at least two oscillator shells.
This renders the shell-model calculation
computationally challenging (if not impossible),
certainly in the heavier nuclei
which are considered here.
We have therefore chosen to test the implied correlations
in the context of a simpler approach,
namely the interacting boson model (IBM) of atomic nuclei~\cite{Iachello87}.
In this model low-lying collective excitations of nuclei
are described in terms of $N_{\rm b}$ bosons
distributed over an $s$ and a $d$ (and sometimes a $g$) level
which can be thought of as correlated pairs of nucleons
occupying valence shell-model orbits
coupled to angular momentum zero and two (and four), respectively.
The number of bosons $N_{\rm b}$
is thus half the number of nucleons in the valence shell.

As the charge radius operator in the IBM
we take the following scalar expression
in terms of the algebra's [U(6) or U(15)] generators~\cite{Iachello87,note1}:
\begin{equation}
\hat T(r^2)=\langle r^2\rangle_{\rm c}+\alpha N_{\rm b}
+\frac{1}{N_{\rm b}}\left(\eta\,\hat n_d+\gamma\,\hat n_g\right),
\label{e_r2ibm}
\end{equation}
where $\langle r^2\rangle_{\rm c}$ is the charge radius of the core nucleus
and $\hat n_d$ ($\hat n_g$) is the $d$($g$)-boson number operator;
$\alpha$, $\eta$, and $\gamma$ are parameters with units of length$^2$.
Then, in analogy with Eq.~(\ref{e_e0r2}),
the appropriate form of the E0 transition operator is
\begin{equation}
\hat T({\rm E0})=
\frac{e_{\rm n}N+e_{\rm p}Z}{N_{\rm b}}
\left(\eta\,\hat n_d+\gamma\,\hat n_g\right).
\label{e_e0ibm}
\end{equation}
Note that for E0 transitions the initial and final states are different
and neither the constant $\langle r^2\rangle_{\rm c}$ nor $N_{\rm b}$
contribute to the transition, so they can be omitted from the E0 operator.
The terms $\hat n_d$ and $\hat n_g$  in Eq.~(\ref{e_r2ibm})
stand for the contribution
from the quadrupole and hexadecapole deformations
to the nuclear radius.
In a first approximation the term in $\hat n_g$
will be omitted from Eqs.~(\ref{e_r2ibm},\ref{e_e0ibm}).
Subsequently, the influence of the $g$ boson
will be explored in $sdg$-IBM.

Although the coefficients $\alpha$ and $\eta$
are treated as parameters and fitted to data on radii,
it is important to understand their physical relevance.
The second term in Eq.~(\ref{e_r2ibm})
depends linearly on particle number
and for a not too large range of nuclei
it can be associated with the isotope shift
originating from the average nuclear charge distribution which varies as
$\langle r^2\rangle_{\rm av}\approx3r_0^2A^{2/3}/5$~\cite{BM},
with $r_0=1.2$~fm.
An estimate of $\alpha$ follows from
\begin{equation}
|\alpha|\approx
\frac{3}{5}r_0^2\left((A+2)^{2/3}-A^{2/3}\right)\approx
\frac{4}{5}r_0^2A^{-1/3},
\label{e_aest}
\end{equation}
which for the nuclei considered here ($A\sim150$)
gives $|\alpha|\sim0.2$~fm$^2$.
The contribution of the quadrupole deformation to the radius
can be estimated as
$\langle r^2\rangle_{\rm def}\approx
5\beta^2\langle r^2\rangle_{\rm av}/4\pi$,
where $\beta$ is the quadrupole deformation parameter
of the geometric model~\cite{BM}.
An estimate of $\eta$ can be obtained
by associating $\langle r^2\rangle_{\rm def}$
with the expectation value $\eta\langle\hat n_d\rangle/N_{\rm b}$
in the ground state of $sd$-IBM.
In a coherent-state approximation
this leads to the relation
\begin{equation}
\eta\frac{{\bar\beta}^2}{1+{\bar\beta}^2}\approx
\frac{4}{3}r_0^2N_{\rm b}^2A^{-4/3}{\bar\beta}^2,
\label{e_best}
\end{equation}
where use has been made of the approximate correspondence
$\beta\approx(4N_{\rm b}/3A)\sqrt{\pi}\bar\beta$
between the quadrupole deformations $\beta$ and $\bar\beta$
in the geometric model and in the IBM, respectively~\cite{Ginocchio80}.
For typical values of $N_{\rm b}\sim10$ and $A\sim150$
this gives a range of possible $\eta$ values between 0.25 and 0.75~fm$^2$.
The preceding analysis also reveals the dependence of $\eta$
on the ratio of the valence to total number of nucleons,
which is expected due the valence character of the IBM.

To test the correlation implied by~(\ref{e_e0r2}),
we have carried out a systematic analysis of even-even nuclei
in the rare-earth region from $Z=58$ to $Z=74$.
Isotope series in this region vary from spherical to deformed shapes,
displaying a more or less sudden shape phase transition.
Such nuclear behavior can be parametrized
in terms of the standard $sd$-IBM Hamiltonian~\cite{Iachello87}.
The details of this calculation
will be reported in a longer publication~\cite{Zerguineun}.
Suffice it to say here that the procedure
is closely related to the one followed
by Garc\'\i a-Ramos {\it et al.}~\cite{Garcia03}
and yields a root-mean square deviation for an entire isotopic chain
which is typically of the order of 100~keV.
In deformed nuclei one adjusts the IBM Hamiltonian
to the observed ground-state band
and $\beta$- and $\gamma$-vibrational bands, if known.
Care should be taken, however, with the identification of the $\beta$-vibrational band.
Our criterion has been to select the band with the strongest $\rho^2$
(not necessarily the first-excited $K^\pi=0^+$ band)
since this is one of the characteristic features
of a $\beta$-vibrational band~\cite{Reiner61}.

\begin{figure}
\includegraphics[width=6.5cm]{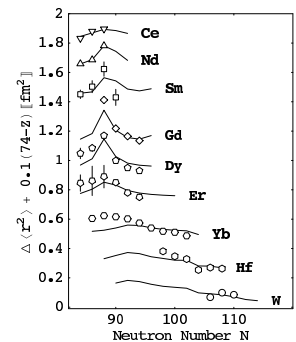}
\caption{
\label{f_shift}
Observed (symbols) and calculated (lines)
isotope shifts (in fm$^2$)
for isotopic chains in the rare-earth region.
The quantity $0.1(74-Z)$~fm$^2$
is added to $\Delta\langle r^2\rangle$ for display purposes.
Data are from Ref.~\cite{isotope}.}
\end{figure}

With wave functions fixed from the energy spectrum
one can now compute nuclear radii using the operator~(\ref{e_r2ibm}).
In the study of nuclear phase-transitional behavior
it is more relevant to plot,
instead of the radii $\langle r^2\rangle$ themselves,
the isotope shifts $\Delta\langle r^2\rangle^{(A)}\equiv
\langle r^2\rangle^{(A+2)}-\langle r^2\rangle^{(A)}$.
The parameter $|\alpha|$ is adjusted to the different isotopic chains
($|\alpha|=0.22$, 0.24, 0.26, 0.13, 0.15, 0.15, 0.11, 0.10, and 0.11~fm$^2$
for even $Z$ between 58 and 74)
but otherwise all calculated isotope shifts in Fig.~\ref{f_shift}
are obtained with a single parameter $\eta=0.5$~fm$^2$.
This value is rather reliably determined
from the peaks in the isotope shifts in the spherical-deformed transitional region
and reproduces the essential features of charge radii in all isotopic chains
with the exception of the light Yb isotopes.

A more direct way to fix the parameter $\eta$
is from isomer shifts
since this quantity is directly proportional to $\eta$.
Of the few isomer shifts known in the rare-earth region,
only the Gd isomer shifts have been measured
with different techniques to give consistent values
and these agree with the theoretical results calculated
with $\eta=0.5$~fm$^2$~\cite{Zerguineun}.

With the parameter $\eta$ determined from isotope and isomer shifts,
one can now compute $\rho^2$ values in the rare-earth region
using the E0 transition operator~(\ref{e_e0ibm}).
The results are compared with the available data in Table~\ref{t_rho}.
\begin{table}
\caption{
\label{t_rho}
Calculated and experimental~\cite{Wood99,Kibedi05} $\rho^2$ values
between levels with angular momentum $J$
and initial and final energies $E_{\rm i}$ and $E_{\rm f}$ (in keV).}
\begin{ruledtabular}
\begin{tabular}{rcrcrcccccl}
Isotope&&\multicolumn{3}{c}{Transition}&&$J$&&\multicolumn{3}{c}{$\rho^2\times10^3$}\\
\cline{3-5}\cline{9-11}
&&$E_{\rm i}$&&$E_{\rm f}$&&&&Calc&&~~~Expt\\
\hline
$^{150}$Sm&&  740&$\rightarrow$&    0&&0&&\phantom{0}8&&\phantom{0.}18~{\sl2}\\
                    &&1046&$\rightarrow$&334&&2&&18&&\phantom{.}100~{\sl40}\\
$^{152}$Sm&&  685&$\rightarrow$&    0&&0&&57&&\phantom{0.}51~{\sl5}\\
                   &&   811&$\rightarrow$&122&&2&&45&&\phantom{0.}69~{\sl6}\\
                   &&1023&$\rightarrow$&366&&4&&32&&\phantom{0.}88~{\sl14}\\
                   &&1083&$\rightarrow$&    0&&0&&\phantom{0}3&&\phantom{0}0.7~{\sl4}\\
                   &&1083&$\rightarrow$&685&&0&&52&&\phantom{0.}22~{\sl9}\\
$^{152}$Gd&&  615&$\rightarrow$&    0&&0&&76&&\phantom{0.}63~{\sl14}\\
                   &&  931&$\rightarrow$&344&&2&&86&&\phantom{0.}35~{\sl3}\\
$^{154}$Gd&&  681&$\rightarrow$&    0&&0&&95&&\phantom{0.}89~{\sl17}\\
                  &&   815&$\rightarrow$&123&&2&&74&&\phantom{0.}74~{\sl9}\\
$^{156}$Gd&&1049&$\rightarrow$&    0&&0&&58&&\phantom{0.}42~{\sl20}\\
                   &&1129&$\rightarrow$&  89&&2&&46&&\phantom{0.}55~{\sl5}\\
$^{158}$Gd&&1452&$\rightarrow$&    0&&0&&34&&\phantom{0.}35~{\sl12}\\
                   &&1517&$\rightarrow$&  79&&2&&30&&\phantom{0.}17~{\sl3}\\
$^{158}$Dy&&1086&$\rightarrow$&   99&&2&&47&&\phantom{0.}27~{\sl12}\\
$^{160}$Dy&&1350&$\rightarrow$&   87&&2&&32&&\phantom{0.}17~{\sl4}\\
$^{162}$Er&&1171&$\rightarrow$& 102&&2&&43&&\phantom{.}630~{\sl460}\\
$^{164}$Er&&1484&$\rightarrow$&   91&&2&&29&&\phantom{0.}90~{\sl50}\\
$^{166}$Er&&1460&$\rightarrow$&     0&&0&&15&&\phantom{0.0}2~{\sl1}\\
$^{170}$Yb&&1229&$\rightarrow$&    0&&0&&36&&\phantom{0.}27~{\sl5}\\
$^{172}$Yb&&1405&$\rightarrow$&    0&&0&&34&&0.20~{\sl3}\\
$^{174}$Hf&&  900&$\rightarrow$&   91&&2&&36&&\phantom{0.}27~{\sl13}\\
$^{176}$Hf&&1227&$\rightarrow$&   89&&2&&17&&\phantom{0.}52~{\sl9}\\
$^{178}$Hf&&1496&$\rightarrow$&   93&&2&&36&&\phantom{0.}14~{\sl3}\\
$^{182}$W&&1257&$\rightarrow$& 100&&2&&51&&\phantom{0}3.5~{\sl3}\\
$^{184}$W&&1121&$\rightarrow$& 111&&2&&58&&\phantom{0}2.6~{\sl5}\\
\end{tabular}
\end{ruledtabular}
\end{table}

Table~\ref{t_rho} illustrates the successes and failures of the present approach.
In the Sm, Gd, and Dy isotopes,
the model reproduces the correct order of magnitude of the $\rho^2$ values 
for a reasonable choice of effective charges, $e_{\rm n}=0.5e$ and $e_{\rm p}=e$.
In the heavier nuclei, however, one observes some glaring discrepancies,
most notably in $^{166}$Er, $^{172}$Yb, and $^{182-184}$W.
One possible reason is that the strong $\rho^2$ to the $\beta$-vibrational band
has not yet been identified experimentally in these nuclei.
This seems to be the case in $^{166}$Er
where recently Wimmer {\it et al.}~\cite{Wimmerun}
measured a $\rho^2$ value of 127~(60)~$10^{-3}$
to an excited $0^+$ state at 1934~keV.
At present this observation is difficult to place
in the systematics of the $\beta$-vibrational band in the Er isotopes
and therefore a re-measurement of E0 properties seems in order
before attempting a re-interpretation of these nuclei. 
Many $\rho^2$ values have been measured in $^{172}$Yb
but none of them is large leading to a confusing situation
that also deserves to be revisited experimentally.
Only in the W isotopes it seems certain that the observed E0 strength
is consistently an order of magnitude smaller than the calculation.
It is known that these nuclei
are in a region of hexadecapole deformation~\cite{Lee74}.
This may offer a qualitative explanation of the suppression of the E0 strength,
as is argued below.

From the preceding analysis the following picture emerges.
Isotopic chains in the rare-earth region exhibit a spherical-to-deformed evolution
which, at the phase-transitional point, is characterized by a peak in the isotope shifts.
Parallel with this behavior there should be a peak in the E0 strength
from the ground to an excited $0^+$ state.
As emphasized by von Brentano {\it et al.}~\cite{Brentano04},
an inescapable prediction of the $sd$-IBM
is that sizable E0 strength should be observed in {\em all} deformed nuclei.
We now investigate to what extent this conclusion is `robust'
by studying a well-known extension of $sd$-IBM
through the introduction of a $g$ boson.

For a review of studies with the $sdg$-IBM
we refer the reader to Devi and Kota~\cite{Devi92}.
A spherical-to-deformed transition
occurs between the limits ${\rm U}(5)\otimes{\rm U}(9)$
and SU(3) of the $sdg$-U(15) model~\cite{Devi90}.
Up to a scale factor, irrelevant for the subsequent discussion,
a schematic Hamiltonian, transitional between the two limits,
is of the form
\begin{equation}
\hat H=(1-\zeta)(\hat n_d+\lambda\hat n_g)-\frac{\zeta}{4N_{\rm b}}\hat Q\cdot\hat Q,
\label{e_sdgh}
\end{equation}
where $\hat Q_\mu$ is the SU(3) quadrupole operator
of the $sdg$-IBM~\cite{Kota87}.
The ${\rm U}(5)\otimes{\rm U}(9)$ limit is obtained for $\zeta=0$
whereas the SU(3) limit corresponds to $\zeta=1$.
By varying $\zeta$ from 0 to 1 one will thus cross the critical point $\zeta_{\rm c}\approx0.5$
at which the spherical-to-deformed transition occurs.

Analytic expressions can be derived
for the matrix elements of the operators $\hat n_s$, $\hat n_d$, and $\hat n_g$
for the limiting values of $\zeta$ in the Hamiltonian~(\ref{e_sdgh})~\cite{Zerguineun}.
It is instructive to compare these results to the corresponding ones in $sd$-IBM
which is done in Table~\ref{t_clme}
in the classical limit $N_{\rm b}\rightarrow\infty$ of SU(3).
\begin{table}
\caption{
\label{t_clme}
Matrix elements in the classical limit of SU(3) in $sd$-IBM and $sdg$-IBM.}
\begin{ruledtabular}
\begin{tabular}{c|ccccccccccc}
&\multicolumn{5}{c}{$\langle0^+_1|\hat n_l|0^+_1\rangle$}
&&\multicolumn{5}{c}{$\langle0^+_1|\hat n_l|0^+_2\rangle$}\\
\cline{2-6}\cline{8-12}
IBM&$l=0$&&$l=2$&&$l=4$&~&$l=0$&&$l=2$&&$l=4$\\
\hline
$sd$&
$\displaystyle{\frac{1}{3}N_{\rm b}}$&&
$\displaystyle{\frac{2}{3}N_{\rm b}}$&&---&&
$\displaystyle{\frac{2}{3}\sqrt{\frac{N_{\rm b}}{2}}}$&&
$\displaystyle{-\frac{2}{3}\sqrt{\frac{N_{\rm b}}{2}}}$&&---\\[2ex]
$sdg$&
$\displaystyle{\frac{1}{5}N_{\rm b}}$&&
$\displaystyle{\frac{4}{7}N_{\rm b}}$&&
$\displaystyle{\frac{8}{35}N_{\rm b}}$&&
$\displaystyle{\frac{2}{5}\sqrt{\frac{N_{\rm b}}{3}}}$&&
$\displaystyle{\frac{2}{7}\sqrt{\frac{N_{\rm b}}{3}}}$&&
$\displaystyle{-\frac{24}{35}\sqrt{\frac{N_{\rm b}}{3}}}$\\
\end{tabular}
\end{ruledtabular}
\end{table}
Considering first the expectation values of $\hat n_l$ in the ground state,
we note that $d$ bosons are dominant in $0^+_1$ both in $sd$- and $sdg$-IBM
and that the contribution of $g$ bosons in $sdg$-IBM is small.
One therefore does not expect a significant effect
of the $g$ boson on the nuclear radius,
and this should be even more so away from the SU(3) limit
for a realistic choice of boson energies, $0<\epsilon_d<\epsilon_g$.
Turning to the $\langle0^+_1|\hat n_l|0^+_2\rangle$ matrix elements,
we see from Table~\ref{t_clme}
that the matrix elements of $\hat n_d$ and $\hat n_g$ in $sdg$-IBM
are of comparable size and different sign.
Therefore, while changes in the nuclear radius
due to the $g$ boson are expected to be small,
one cannot rule out its significant impact
on the $\rho^2(0^+_\beta\rightarrow0^+_1)$ values in deformed nuclei.

\begin{figure}
\includegraphics[width=4.25cm]{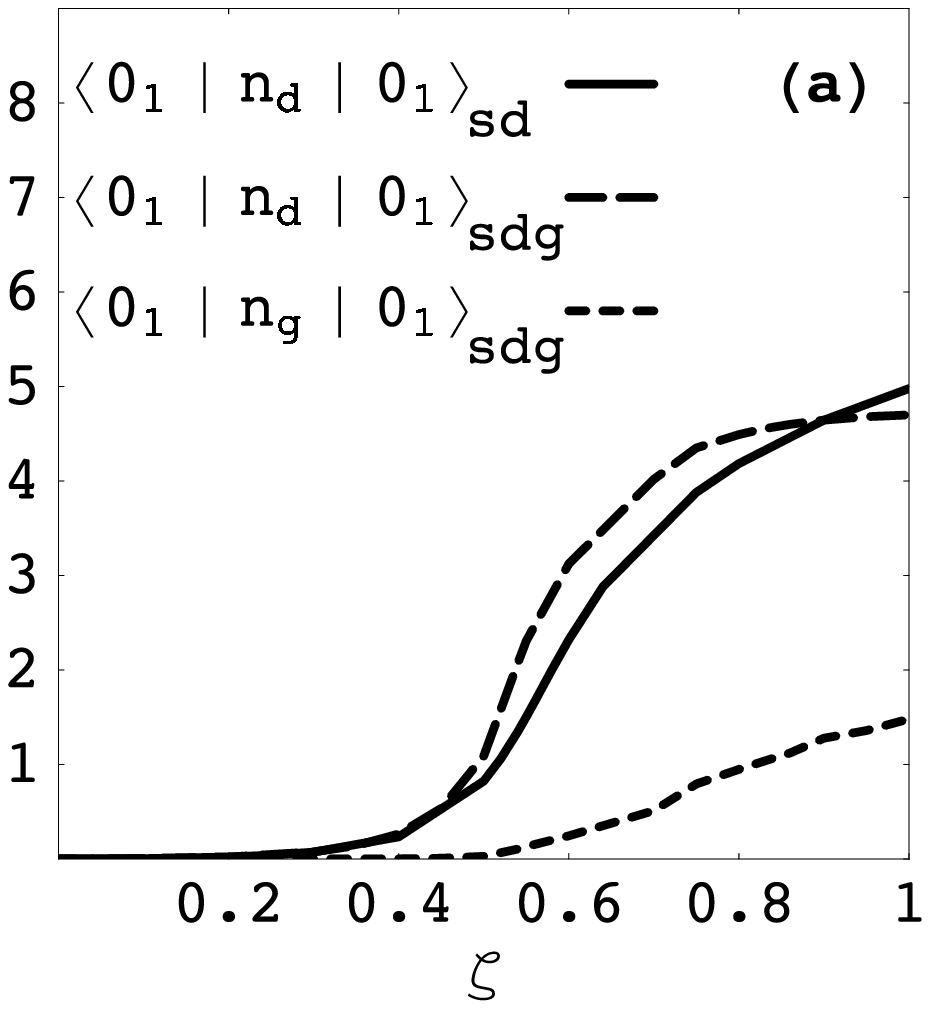}
\includegraphics[width=4.25cm]{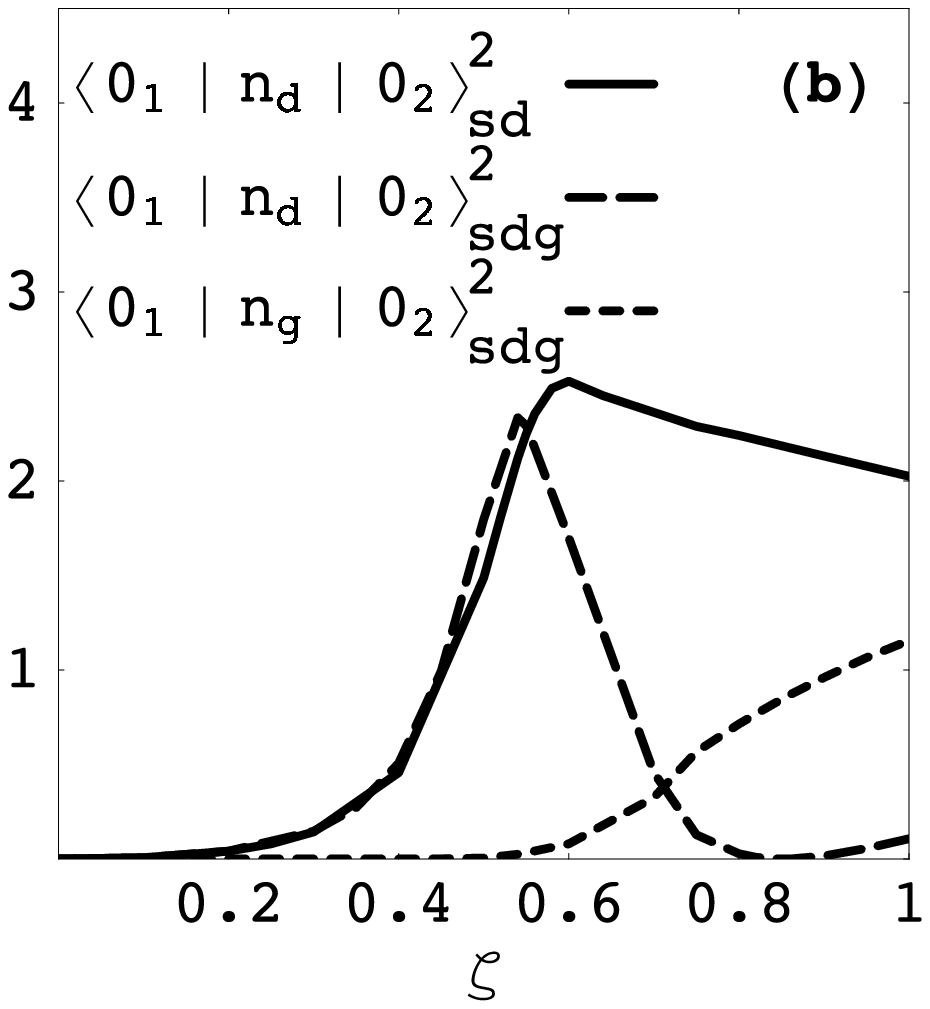}
\caption{
\label{f_ndg}
The matrix elements (a) $\langle0^+_1|\hat n_l|0^+_1\rangle$
and (b) $\langle0^+_1|\hat n_l|0^+_2\rangle^2$ for $l=2$ and $l=4$
in the spherical-to-deformed transition of $sd$-IBM and $sdg$-IBM.
In $sd$-IBM the transition is from U(5) to SU(3)
and $sdg$-IBM from ${\rm U}(5)\otimes{\rm U}(9)$ to SU(3) with $\lambda=1.5$.
The number of bosons is $N_{\rm b}=8$.}
\end{figure}

This argument can be made more quantitative
by studying the spherical-to-deformed shape transition of the Hamiltonian~(\ref{e_sdgh}).
Using a numerical code~\cite{Heinzeun} the matrix elements of $\hat n_d$ and $\hat n_g$
can be calculated for arbitrary $\zeta$.
For the ratio of boson energies we choose $\lambda=1.5$.
The results are shown in Fig.~\ref{f_ndg}
and compared with the matrix elements of $\hat n_d$
calculated for the U(5)-to-SU(3) transition in $sd$-IBM.
Panel (a) of the figure confirms the dominance
of the $d$ boson in the ground state of deformed nuclei
both in $sd$- and $sdg$-IBM.
Moreover, the expectation value of $\hat n_d$ varies with $\zeta$
in very much the same way in both models.
In $sd$-IBM as well as in $sdg$-IBM
the sharp increase in $\langle0^+_1|\hat n_d|0^+_2\rangle^2$
is observed around $\zeta_{\rm c}\approx0.5$.
Up to that point there is essentially no contribution
to $\rho^2(0^+_2\rightarrow0^+_1)$ from the $g$ boson.
Consequently, all $sd$-IBM E0 results up to the phase-transitional point
are not modified significantly by the $g$ boson.
As can be seen from Fig.~\ref{f_ndg}b,
in the deformed regime this is no longer true
since, in $sdg$-IBM, a sharp decrease of $\langle0^+_1|\hat n_d|0^+_2\rangle^2$
occurs at $\zeta\approx0.5$
and $\langle0^+_1|\hat n_g|0^+_2\rangle^2$
rapidly increases after $\zeta\approx0.6$
and dominates $\langle0^+_1|\hat n_d|0^+_2\rangle^2$ for $\zeta\geq0.7$.

In summary, the correlation between radii and E0 transitions
has been investigated in transitional nuclei.
A quantitative analysis of radii and $\rho^2$ values in the rare-earth nuclei
seems to validate the explanation of E0 strength
which is based on a geometric picture of the nucleus
as advocated a long time ago~\cite{Reiner61}.
The proposed correlation depends on the effective charges
appearing in the radius and E0 operators, assumed to be the same,
and establishes a method to determine these charges empirically.
The hexadecapole deformation
which has only a weak effect on the nuclear radius
but possibly a strong one on E0 transitions,
might perturb the correlation in deformed nuclei.

We wish to thank Ani Aprahamian, Lex Dieperink, and Kris Heyde for helpful discussions.
This work has been carried out in the framework of CNRS/DEF project N 19848.
S.Z.\ thanks the Algerian Ministry of High Education and Scientific Research for financial support.



\begin{references}
\bibitem{note1}
The factor $1/N_{\rm b}$
is included because it is the {\em fraction} $\langle\hat n_d\rangle/N_{\rm b}$
which is a measure of the quadrupole deformation.
Slightly less good but still reasonable results
are obtained  without the factor $1/N_{\rm b}$.
\end{references}
\end{document}